\newcommand{\be}{\begin{eqnarray}}
\newcommand{\ee}{\end{eqnarray}}

\newcommand{\para}{||}
\documentclass[twocolumn,aps,showpacs]{revtex4}
\usepackage{graphics}
\begin{document}
\title{Effective potential at finite temperature in a constant
magnetic field I: Ring diagrams in a scalar theory} 
\author{Alejandro Ayala,$^\dagger$ Angel S\'anchez$^\dagger$, Gabriella
        Piccinelli$^\ddagger$, Sarira Sahu$^\dagger$}  
\affiliation{$\dagger$Instituto de Ciencias Nucleares, Universidad Nacional 
         Aut\'onoma de M\'exico, Apartado Postal 70-543, 
         M\'exico Distrito Federal 04510, M\'exico.\\
         $^\ddagger$Centro Tecnol\'ogico, ENEP Arag\'on,
         Universidad Nacional Aut\'onoma de M\'exico,
         Avenida Rancho Seco S/N, Bosques de Arag\'on,
         Nezahualc\'oyotl, Estado de 
         M\'exico 57130, M\'exico.}
\begin{abstract}
We study symmetry restoration at finite temperature in the theory of a
charged scalar field interacting with a constant, external magnetic
field. We compute the finite temperature effective potential
including the contribution from ring diagrams. We show that in 
the weak field case, the presence of the field produces a stronger
first order phase transition and that the temperature for the onset of
the transition is lower, as compared to the case without magnetic
field. 
\end{abstract}


\maketitle

\section{Introduction}\label{I}

Symmetry restoration in field theories at finite temperature has been
a subject of interest for quite some time already, in particular when
applied to the description of phase transitions in the early
universe. An important example is the study of the nature of the
electroweak phase transition (EWPT) in the standard model (SM) for
temperatures of order 100 GeV~\cite{reviewsEWPT}. It is by now well
known that the 
correct description of this phase transition requires accounting for
non-perturbative phenomena casted in terms of the so called {\it ring
diagrams}. Inclusion of this terms has the important effect of
changing the nature of the phase transition from second to first order.

In recent years it has also become important to study the influence that
magnetic fields could have had on cosmological phase
transitions~\cite{reviews}. Though the
nature and origin of these fields is unknown it is certainly true that
the current limits on their strength during the EWPT cannot
rule them out.

Possible consequences for the propagation of fermions during a
first order EWPT in the presence of magnetic fields such as the
generation of an axial asymmetry~\cite{Ayala} or a spin-up spin-down
asymmetry~\cite{Campanelli} have been recently studied.
On the other hand, it has been shown that magnetic fields are also
able to generate a stronger first order EWPT as compared to the case
when these fields are not
present~\cite{{Giovannini},{Elmfors},{Giovannini2}}. Nevertheless, 
these studies are either classical or resort to perturbation theory to
lowest order.  In contrast to these perturbative estimates, lattice 
calculations~\cite{Kajantie} seem to indicate that, for Higgs masses
$m_H \ge 80$ GeV, the presence of a magnetic field does not suffice to
make the transition to be of first order. 
In this context, the question emerges as to what is the
effect of a magnetic field in the description of the phase transition
when also including the contribution of non-perturbative effects such
as the ring diagrams at finite temperature. 

To our knowledge, only one attempt in this direction has been
made. This is Ref.~\cite{Skalozub2} where this question is addressed
in the context of the generation of baryon number in the SM during the
EWPT. Unfortunately, neither the details nor the limitations of the
approximations involved are stated and thus the need for a closer look
at this phenomenon. 

Recall that field theoretical calculations involving external magnetic
fields can be carried out by means of Schwinger's proper time
method~\cite{Schwinger}. The method incorporates to all orders the
effects of the external field into the Green's functions of the
theory. To manage the expressions thus obtained, it is
customary to resort to either the strong or the weak field limits. For
theories involving particles with mass --as is the case of
theories with spontaneous breaking of the symmetry-- and at finite
temperature, it is therefore mandatory to clearly state the hierarchy
of the three energy scales involved when carrying out the approximations.

In this work, we study the problem of symmetry restoration at finite
temperature in the presence of an external magnetic field. We compute
the finite temperature effective potential, up to the contribution of
ring diagrams, for a charged scalar field interacting with a uniform
external magnetic field. We point out that the problem is not merely
of academic interest since similar arguments apply to the case of the
SM degrees of freedom. However, the complexity of expressions of an
exact method makes it necessary to work first in a simpler, though
relevant case, to have a better control over the approximations and
results and latter on extend them to scenarios where more degrees of
freedom are involved. 

The work is organized as follows: In Sec.~\ref{secII} we find the
propagator for the charged scalar field in the presence of an external
magnetic field. From the exact expression we compute the weak and
strong field limits of this propagator. In Sec.~\ref{secIII} we work
out the finite temperature effective potential up to the contribution
of the ring diagrams, also in the weak and strong field limits. In
Sec.~\ref{secIV} we use the expressions for the effective potential to
discuss symmetry restoration. We show that the presence of the
external field makes the phase transition strongly first order in the
weak field limit. We finally conclude in Sec.~\ref{secV}. We reserve
for the appendix the explicit calculation of integrals appearing
throughout the work. 

\section{Scalar propagator in a constant magnetic field}\label{secII}

Using Schwinger's proper-time method, it is possible to obtain the
exact expression for the vacuum propagator for a charged scalar boson
with charge $e$, in the presence of an external magnetic field,
$D^B(x',x'')$, which is given by 
\be
   D^B(x',x'')=
   \varphi(x',x'')\int\frac{d^4k}{(2\pi)^4}e^{-ik\cdot
   (x'-x'')}D^B(k),
   \label{prop}
\ee
where
\be
  iD^B(k)=
   \int_0^\infty\frac{ds}{\cos eBs}
   e^{is\left(k_{\para}^2 - k_\perp^2
   \frac{\tan eBs}{eBs} - m^2
   +i\epsilon \right)}.
   \label{defsD0}
\ee
Similarly, the expression for the scalar boson self-energy
$\Pi^B(x',x'')$ is given by
\be
   \Pi^B(x',x'')=
   \varphi(x',x'')\int\frac{d^4k}{(2\pi)^4}e^{-ik\cdot
   (x'-x'')}\Pi^B(k).
   \label{defPi}
\ee
In Eqs.~(\ref{defsD0}) and~(\ref{defPi}) we use the definitions
\be
   (a\cdot b)_{\para}&=&a^0b^0-a^3b^3\nonumber\\
   (a\cdot b)_{\perp}&=&a^1b^1+a^2b^2\, ,
\ee
for any two four vectors $a^\mu$, $b^\mu$. The phase factor $\varphi$ in
Eq.~(\ref{prop}) is given by 
\be
   \varphi(x',x'')=\exp\left[ie\int_{x''}^{x'}dx_\mu A^\mu 
   (x)\right]\, .
   \label{phase}
\ee
and does not depend on the integration path. 
Since from now on we will be concerned with expressions such the
one-loop self-energy or the effective potential that do not have a
momentum dependence, and are thus diagonal in coordinate space, the
phase factor of Eq.~(\ref{phase}) vanishes and it will be enough to
work in the momentum representation.

Notice that taking the limit $eB\rightarrow 0$ in Eq.~(\ref{defsD0})
and by means of the identity 
\be
   \frac{1}{q^2+i\epsilon}=-i
   \int_{0}^\infty ds e^{is(q^2+i\epsilon )}\, ,
   \label{rep}
\ee 
one obtains the free Feynman vacuum propagator for the scalar field
given by 
\be
  i D^F(k)=\frac{i}{k^2-m^2 + i\epsilon}.
   \label{freeprop}
\ee

\begin{figure}[t!] 
\vspace{0.4cm}
{\centering
\resizebox*{0.3\textwidth}
{0.15\textheight}{\includegraphics{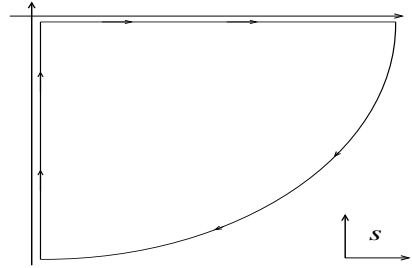}}
\par}
\caption{Integration contour in the complex $s$-plane to compute the
integral representing the scalar propagator in the presence of a
magnetic field.}
\label{fig1}
\end{figure}

Let us now proceed to work out Eq.~(\ref{defsD0}) to find a
working representation for the scalar propagator $iD^B(k)$. First we
do the change of variable $eBs\rightarrow s$ to write
Eq.~(\ref{defsD0}) as
\be
   iD^B(k)=\frac{1}{eB}
   \int_0^\infty\frac{ds}{\cos s}
   e^{i\frac{s}{eB}\left(k_{\para}^2 - k_\perp^2
   \frac{\tan s}{s} - m^2
   +i\epsilon \right)}.
   \label{defsD0s}
\ee
The integrand in Eq.~(\ref{defsD0s}) is analytical in the
lower complex $s$-plane and the zeros of $\cos (s)$ are all located
on the real $s$-axis. Furthermore, the $i\epsilon$ in the exponent
ensures that for $|s|\rightarrow \infty$, the integrand dies out
sufficiently rapidly. Therefore we can close the contour of
integration on a path whose first leg is a horizontal line just below
the real $s$-axis, continued along the 
quarter-circle at infinity in the right-lower quadrant and finally
along the negative imaginary $s$-axis. This is depicted in
Fig.~\ref{fig1}. Using Cauchy's theorem, the
integral in Eq.~(\ref{defsD0s}) can be written as
\be
   iD^B(k)=\frac{-1}{eB}
   \int_{-i\infty}^0\frac{ds}{\cos s}
   e^{i\frac{s}{eB}\left(k_{\para}^2 - k_\perp^2
   \frac{\tan s}{s} - m^2
   +i\epsilon \right)}.
   \label{defsD0smod}
\ee
Since the integration in Eq.~(\ref{defsD0smod}) is along the imaginary
axis, we make the change of variable $s=-i\tau$ with $\tau$ real and
thus Eq.~(\ref{defsD0smod}) becomes
\be
   iD^B(k)=\frac{-i}{eB}
   \int_0^\infty\!\!\!\!\frac{d\tau}{\cos (-i\tau)}
   e^{\frac{\tau}{eB}\left(k_{\para}^2 - k_\perp^2
   \frac{\tan (-i\tau)}{(-i\tau)} - m^2
   +i\epsilon \right)}\!\!.
   \label{defsD0smodtau}
\ee
Notice that since $\tau \geq 0$, this last integral converges for
Re$(k_{\para}^2 + i\epsilon) <0$, that is $k_0^2-k_3^2 <0$ which means
that we are considering momenta in Eucledian space. Though the result
can later on be analytically continued to Minkowski space, we will
continue considering $k^\mu$ in Euclidean space and for finite
temperature calculations we will work in the imaginary-time formalism.

Next, we use that
\be
   \cos (-i\tau) &=& \frac{e^\tau + e^{-\tau}}{2}\nonumber\\
   i\tan (-i\tau) &=& \frac{e^\tau - e^{-\tau}}
   {e^\tau + e^{-\tau}}\, .
   \label{costan}
\ee
Introducing the variable $u=e^{-2\tau}$, we can write
\be
   \frac{1}{\cos (-i\tau)} &=& \frac{2u^{1/2}}{1+u}\nonumber\\
   i\tan (-i\tau) &=& 1- \frac{2u}{1+u}\, .
   \label{costanconu}
\ee
Using Eq.~(\ref{costanconu}), we can write Eq.~(\ref{defsD0smodtau})
as
\be
   iD^B(k)&=&\frac{-2i}{eB}\int_0^\infty
   d\tau e^{\frac{\tau}{eB}(k_{\para}^2-m^2+i\epsilon )}
   e^{-\frac{k_\perp^2}{eB}}\nonumber\\
   &\times&u^{1/2}\frac{e^{\frac{2k_\perp^2}{eB}
   \left(\frac{u}{1+u}\right)}}{1+u}.
   \label{Dconu}
\ee
Equation~(\ref{Dconu}) is now suited to introduce the generating
function for the Laguerre polynomials~\cite{Dolivo} $L_l(x)$, given by
\be
   \frac{e^{-xz/(1-z)}}{1-z} = \sum_{l=0}^\infty L_l(x)z^l\, ,
   \label{genfun}
\ee
from which, interchanging the order of the summation and the
integration, we can write 
\be
   iD^B(k)&=&\frac{-2i}{eB}
   \sum_{l=0}^\infty (-1)^lL_l\left(\frac{2k_\perp^2}{eB}\right)
   e^{-\frac{k_\perp^2}{eB}}\nonumber\\
   &\times&
   \int_0^\infty d\tau u^{l+1/2}
   e^{\frac{\tau}{eB}(k_{\para}^2-m^2+i\epsilon )}\, .
   \label{Dconsum}
\ee
The integral over $\tau$ can now be explicitly evaluated with the
result
\be
   &&\int_0^\infty\!\!\!\! d\tau
   e^{\frac{\tau}{eB}(k_{\para}^2-(2l+1)eB-m^2+i\epsilon )}=\nonumber\\
   &&\frac{-eB}{k_{\para}^2-(2l+1)eB-m^2+i\epsilon}\, ,
   \label{inttau}
\ee
from which the expression for the propagator finally becomes
\be
   iD^B(k)=2i\sum_{l=0}^\infty
   \frac{(-1)^lL_l\left(\frac{2k_\perp^2}{eB}\right)
   e^{-\frac{k_\perp^2}{eB}}}
   {k_{\para}^2-(2l+1)eB-m^2+i\epsilon}\, .
   \label{Dfin}
\ee

\subsection{Weak field limit}\label{secIIIa}

Let us now work out Eq.~(\ref{Dfin}) in the limit where $eB$ is
small compared to the momenta. For this purpose, we follow
Ref.~\cite{Tzuu} and reorganize the series in Eq.~(\ref{Dfin}) in
powers of $(eB)$ to make evident the lowest contributing power of
$(eB)$ which is the most important one in this limit.
\be
   iD^B(k)&=&2i
   \frac{e^{-\frac{k_\perp^2}{eB}}}{(k_{\para}^2-m^2)}\nonumber\\
   &\times&\sum_{l=0}^\infty
   \frac{(-1)^lL_l\left(\frac{2k_\perp^2}{eB}\right)}
   {1-(2l+1)eB/(k_{\para}^2-m^2)}\, ,
   \label{Dfinprep}
\ee
where in anticipation to working in the imaginary-time formulation of
thermal field theory, we have omitted the $i\epsilon$ term. Notice
that we can formally write
\be
   \frac{1}{1-(2l+1)eB/(k_{\para}^2-m^2)}=\sum_{j=0}^\infty
   \left(\frac{eB[2l+1]}{k_{\para}^2-m^2}\right)^j\, ,
   \label{apros}
\ee
from which the propagator can be written as
\be
   iD^B(k)&\!\!\!=\!\!\!&\frac{i}{(k_{\para}^2-m^2)}\sum_{j=0}^\infty
   \left(\frac{eB}{k_{\para}^2-m^2}\right)^j\nonumber\\
   &\!\!\!\!\!\!&
   \left\{2e^{-\frac{k_\perp^2}{eB}}\sum_{l=0}^\infty
   (-1)^lL_l\left(\frac{2k_\perp^2}{eB}\right)
   (2l+1)^j\right\}\!\! .
  \label{aftersum}
\ee
Notice that Eq.~(\ref{aftersum}) is valid for $eB \ll m^2$. The sum in
the term between curly brackets in Eq.~(\ref{aftersum}), namely
\be
   S_j\equiv\left\{2e^{-\frac{k_\perp^2}{eB}}\sum_{l=0}^\infty
   (-1)^lL_l\left(\frac{2k_\perp^2}{eB}\right)
   (2l+1)^j\right\}\, ,
   \label{curly}
\ee
represents a special case of the identity
\be
   f(x)&\equiv&\frac{e^{-i\left(\frac{k_\perp^2}{eB}\right)
   \tan (x)}}{\cos (x)}\nonumber\\
   &=&2e^{-\frac{k_\perp^2}{eB}}\sum_{l=0}^\infty
   (-1)^lL_l\left(\frac{2k_\perp^2}{eB}\right)
   e^{-i(2l+1)x}\, .
   \label{deffx}
\ee
Therefore, we see that for a given $j$, $S_j$ is given by
\be
   S_j=\left.i^j\frac{d^jf}{dx^j}\right|_{x=0}\, .
   \label{Sj}
\ee
It is now a simple exercise to write down the propagator as a series
in powers of $eB$. Keeping only the lowest order terms, we get 
\be
   iD^B(k)\!\!\!\!\!\!&&\stackrel{eB\rightarrow 0}{\longrightarrow}
   \frac{i}{k_{\para}^2-k_\perp^2-m^2}\nonumber\\
   &&\times
   \left\{1-\frac{(eB)^2}{(k_{\para}^2-k_\perp^2-m^2)^2}
   -\frac{2(eB)^2(k_\perp^2)}{(k_{\para}^2-k_\perp^2-m^2)^3}
   \right\}.\nonumber\\
   \label{firsteB}
\ee
   
\subsection{Strong field limit}\label{secIIIb}

For the purpose of considering the limit where $eB$ is large, recall
that Eq.~(\ref{Dfin}) can be thought of as expressing the scalar
propagator in terms of a superposition of contributions from the {\it
Landau levels}, each of which corresponds to a discrete energy 
given by 
\be
   E_l=\sqrt{k_{3}^2+(2l+1)eB+m^2}\, .
   \label{LL}
\ee
For $eB$ large compared to the momenta, the gap between
successive energy levels, $\Delta E\simeq 2eB$ becomes large. When
working at finite temperature, where the momentum becomes an energy
scale of order $T$, thermal fluctuations will rarely produce
occupation of excited energy levels and thus, for $eB\gg T$, it is a
good approximation to consider only the contribution from the lowest
Landau level and write
\be
   iD^B(k)\stackrel{eB\rightarrow\infty}{\longrightarrow}
   2i\ \frac{e^{-\frac{k_\perp^2}{eB}}}{k_{\para}^2-eB-m^2}\, .
   \label{LLL}
\ee
In this approximation, transverse and longitudinal momenta decouple. 

In what follows, we will work either with Eq.~(\ref{firsteB}) or
Eq.~(\ref{LLL}) when discussing the effective potential in the weak
or strong field limits, respectively.

\begin{figure}[t!] 
\vspace{0.4cm}
{\centering
\resizebox*{0.125\textwidth}
{0.1\textheight}{\includegraphics{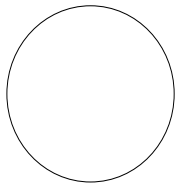}}
\par}
\caption{Feynman diagram representing the one-loop vacuum bubble
contribution to the effective potential in the absence of the magnetic
field.} 
\label{fig2}
\end{figure}

\section{Effective potential}\label{secIII}

To include quantum corrections to the tree level potential, we recall
that it is convenient to express these as a series in powers of the
coupling constant $\lambda$. In what follows, we work in the
imaginary-time formulation of thermal field theory. First, we consider
that the integration over four momenta is carried out in Eucledian
space with $k_0=ik_4$. This means that 
\be
   \int\frac{d^4k}{(2\pi)^4}&=&i\int\frac{d^4k_E}{(2\pi)^4}\, .
   \label{ome}
\ee
Next, we recall that in the formalism, energies take discrete
values, namely $k_4=w_n=2n\pi T$ with $n$ an integer as corresponds to
a Matsubara frequency for bosons and thus
\be
   \int\frac{d^4k_E}{(2\pi)^4}\rightarrow
   T\sum_n\int\frac{d^3k}{(2\pi)^3}\, . 
   \label{fromMtoE}
\ee
In this manner, the one-loop contribution to the effective,
finite-temperature potential, whose Feynman dyagram is depicted in
Fig.~\ref{fig2}, is given by~\cite{LeBellac} 
\be
   V^{(1)}&=&\frac{T}{2}\sum_n\int\frac{d^3k}{(2\pi)^3}\ln
   [\Delta^B (k)]^{-1}\, ,
   \label{V1}
\ee
where
\be
   \Delta^B (k)=-D^B(k_0=i\omega_n,{\mathbf{k}})\, .
   \label{delta}
\ee
\begin{figure}[t!] 
\vspace{0.4cm}
{\centering
\resizebox*{0.44\textwidth}
{0.11\textheight}{\includegraphics{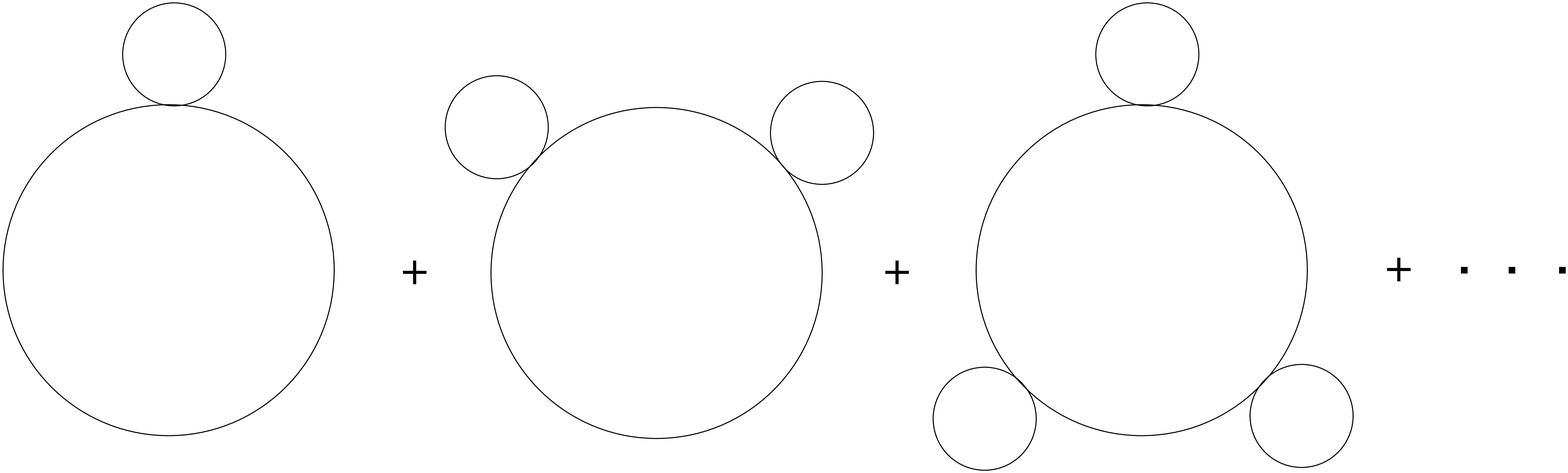}}
\par}
\caption{Feynman diagram representing the ring diagrams
contribution to the effective potential in the absence of the magnetic
field.} 
\label{fig3}
\end{figure}

It is well known, for the case of vanishing external magnetic field,
that the next order correction to Eq.~(\ref{V1}) comes from the so
called {\it ring diagrams}~\cite{Carrington} depicted in
Fig.~\ref{fig3}. As we will show, 
this is  also the case in the presence of an external magnetic
field where the scalar propagator and
self-energy used in the calculation include the effects of the magnetic
field. The contribution to the effective potential arises from the
mode with $n=0$ from the expression given by  
\be
   V^{(\mbox{\tiny{ring}})}&=&-\frac{T}{2}\sum_n\int\frac{d^3k}{(2\pi)^3}
   \sum_{N=1}^\infty\frac{1}{N}\left(\Pi^B\Delta^B (k)\right)^N
   \nonumber\\
   &=&\frac{T}{2}\sum_n\int\frac{d^3k}{(2\pi)^3}
   \ln [1+\Pi^B\Delta^B (k)]\, ,
   \label{Vring}
\ee
where $\Pi^B$ has to be computed also
self-consistently~\cite{Carrington}. Since for the discussion of
symmetry restoration we will consider a theory of a charged scalar
with a self-interaction of the form $(\phi^\dagger\phi)^2/4$ (see
Sec.~\ref{secIV}), the explicit expression for $\Pi^B$ is given by
\be
   \Pi^B=\lambda T\sum_n\int\frac{d^3k}{(2\pi)^3}
   \Delta^B (k;m^2\rightarrow m^2+\Pi_1)
   \label{Pi}
\ee
where
\be
   \Pi_1&=&\lambda T\sum_n\int\frac{d^3k}{(2\pi)^3}
   \frac{1}{\omega_n^2+{\mathbf{k}}^2+m^2}\nonumber\\
   &=&\frac{\lambda T^2}{12} + {\mathcal{O}}(m^2)
   \label{Pi1}
\ee
is the one-loop self-energy. 

We now proceed to compute the expressions in
Eqs.~(\ref{V1}),~(\ref{Vring}) and~(\ref{Pi}) in the weak and strong
field limits.
 
\subsection{Weak field limit}\label{secIVa}

Let us first start with the expression for the
self-energy. Using Eqs.~(\ref{firsteB}) and Eq.~(\ref{delta}) into
Eq.~(\ref{Pi}), we have to explicitly evaluate
\begin{widetext}
\be
   \Pi^B=\lambda T\sum_n\int
   \frac{d^3k}{(2\pi)^3}
   \frac{1}{\omega_n^2+{\mathbf{k}^2}+m^2+\Pi_1}
   \left\{1-\frac{(eB)^2}{(\omega_n^2+{\mathbf{k}^2}+m^2+\Pi_1)^2} 
   +\frac{2(eB)^2(k_\perp^2)}{(\omega_n^2+{\mathbf{k}^2}+m^2+\Pi_1)^3}
   \right\}.
   \label{Piexpl}
\ee
\end{widetext}
\begin{figure}[t!] 
\vspace{0.4cm}
{\centering
\resizebox*{0.66\textwidth}
{0.0805\textheight}{\includegraphics{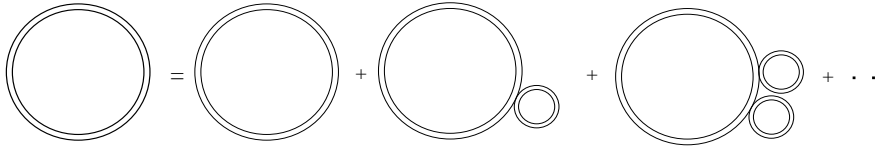}}
\par}
\caption{Feynman diagram representing the effective potential,
including the contribution from the ring diagrams, in the
presence of the magnetic field. The double thin lines represent the
scalar propagator and 
self-energy including the effects of the magnetic field.} 
\label{fig4}
\end{figure}
We will work out Eq.~(\ref{Piexpl}) considering explicitly that the
hierarchy of energy scales is 
\be
   eB\ll m^2\ll T^2\, .
   \label{hirarchy}
\ee
We work in the limit $m \ll T$, since this is the important case for
the contribution from ring diagrams~\cite{LeBellac}. 
The first term in Eq.~(\ref{Piexpl}) corresponds to the finite
temperature $B=0$ contribution. For the hierarchy of energy scales
considered, the leading contribution at finite temperature is
thus~\cite{LeBellac} 
\be
   \Pi^{B=0}&\equiv&\lambda T\sum_n\int
   \frac{d^3k}{(2\pi)^3}
   \frac{1}{\omega_n^2+{\mathbf{k}^2}+m^2+\Pi_1}\nonumber\\
   &=&\frac{\lambda T^2}{12}
   \left\{1-3\left(\frac{\lambda}{12\pi^2}\right)^{1/2}\right\}
   + {\mathcal{O}}(m^2)\, .
   \label{Pi1self}
\ee
Notice that the non-perturbative nature of the resummation method
is signaled by the non-analyticity of the expansion in the coupling
$\lambda$ in Eq.~(\ref{Pi1self}). In order to keep
track of the lowest order corrections in $\lambda$ and to emphasize the
corrections that have to do with the magnetic field, hereafter we
omit the second term in Eq.~(\ref{Pi1self}).

The second term in Eq.~(\ref{Piexpl}) involves the computation of the
integral
\be
   I_1&=&T\int\frac{d^3k}{(2\pi)^3}
   \frac{1}{({\mathbf{k}}^2+m^2+\Pi_1)^3}\nonumber\\  
   &+& T\sum_{n\neq 0}\int\frac{d^3k}{(2\pi)^3}
   \frac{1}{(\omega_n^2+{\mathbf{k}}^2+m^2+\Pi_1)^3}\, ,
   \label{P1A}
\ee
where we have explicitly separated the contribution from the $n=0$
mode from the rest. The first term in Eq.~(\ref{P1A}) is simply
\be
   T\int\frac{d^3k}{(2\pi)^3}\frac{1}{({\mathbf{k}}^2+m^2+\Pi_1)^3}=
   \frac{T}{32\pi (m^2+\Pi_1)^{3/2}}\, .
   \label{firstP1A}
\ee
For the second term, we use the findings of Ref.~\cite{Bedingham} 
which are suited for an expansion for $T^2>m^2+\Pi_1$ with the result
\begin{widetext}
\be
   T\sum_{n\neq 0}\int\frac{d^3k}{(2\pi)^3}
   \frac{1}{(\omega_n^2+{\mathbf{k}}^2+m^2+\Pi_1)^3}=
   \frac{1}{(4\pi)^{3/2}(2\pi)^3}\left(\frac{1}{T^2}\right)
   \sum_{j=0}^{\infty}
   \frac{(-1)^j}{j!}\zeta (2j+3)\Gamma (j+3/2)
   \left(\frac{\sqrt{m^2+\Pi_1}}{2\pi T}\right)^{2j}\, ,
   \label{secondP1A}
\ee
\end{widetext}
where $\zeta$ and $\Gamma$ are the Riemann-zeta function and Gamma
function, respectively. For the hierarchy of energy scales
considered, the leading contribution comes from the mode with $n=0$
and thus
\be
   I_1=\frac{T}{32\pi (m^2+\Pi_1)^{3/2}} +{\mathcal{O}}(1/T^2)\, .
   \label{Aexpl}
\ee
The third term in Eq.~(\ref{Piexpl}) involves the computation of the
integral
\be
   I_2&=&T\sum_{n}\int\frac{d^3k}{(2\pi)^3}
   \frac{k_\perp^2}{(\omega_n^2+{\mathbf{k}}^2+m^2+\Pi_1)^4}\nonumber\\
   &=&T\sum_{n}\int\frac{d^3k}{(2\pi)^3}
   \frac{(2/3){\mathbf{k}}^2}{(\omega_n^2+{\mathbf{k}}^2+m^2+\Pi_1)^4}\, .
   \label{B}
\ee
It is easy to see that for the hierarchy of energy scales
considered, the leading contribution also comes from the mode with
$n=0$ and thus
\be
   I_2=\frac{1}{3}\left(\frac{T}{32\pi (m^2+\Pi_1)^{3/2}}\right)
   +{\mathcal{O}}(1/T^2)\, .
   \label{Bexpl}
\ee
\\
Collecting the results in Eqs.~(\ref{Pi1}),~(\ref{Aexpl}) 
and~(\ref{Bexpl}) into Eq.~(\ref{Piexpl}), the final expression for
the charged scalar self-energy in the weak field limit is given by
\be
   \Pi^B\stackrel{eB\rightarrow 0}{\longrightarrow}
   \frac{\lambda T^2}{12}\left\{1 - \frac{(eB)^2}{8\pi
   T(m^2+\Pi_1)^{3/2}}\right\}\, . 
   \label{weakPi}
\ee 
Notice that the only infinity that appears, and that for the ease of
the discussion we have ignored, corresponds to the usual mass
renormalization at zero temperature.

We now turn to the computation of the effective potential, depicted
in Fig.~\ref{fig4}. To one-loop this is given by
Eq.~(\ref{V1}). Notice that to lowest order in the magnetic field, we
can write
\begin{widetext}
\be
   [\Delta^B (k)]^{-1}\simeq
   (\omega_n^2+{\mathbf{k}}^2+m^2+\Pi_1)
   \Big\{1+\frac{(eB)^2}{(\omega_n^2+{\mathbf{k}}^2+m^2+\Pi_1)^2}
   -\frac{2(eB)^2(k_\perp^2)}{(\omega_n^2+{\mathbf{k}}^2+m^2+\Pi_1)^3}
   \Big\}\, ,
   \label{V1D}
\ee
\end{widetext}
from where we can expand $\ln [\Delta^B (k)]^{-1}$ to lowest order
\begin{widetext}
\be
   \ln [\Delta^B
   (k)]^{-1}\simeq\ln(\omega_n^2+{\mathbf{k}}^2+m^2+\Pi_1)
   +(eB)^2\left\{
   \frac{1}{(\omega_n^2+{\mathbf{k}}^2+m^2+\Pi_1)^2}
   -\frac{2(k_\perp^2)}{(\omega_n^2+{\mathbf{k}}^2+m^2+\Pi_1)^3}
   \right\}
   \label{expln}
\ee
\end{widetext}
Using Eq.~(\ref{expln}), the term at hand becomes
\\
\begin{widetext}
\be
   V^{(1)}=\frac{T}{2}\sum_n\int\frac{d^3k}{(2\pi)^3}\ln
   [\Delta^B (k)]^{-1}&\simeq&
   \frac{T}{2}\sum_n\int\frac{d^3k}{(2\pi)^3}\left\{
   \ln(\omega_n^2+{\mathbf{k}}^2+m^2+\Pi_1)\right.\nonumber\\
   &+&(eB)^2\left.\left[
   \frac{1}{(\omega_n^2+{\mathbf{k}}^2+m^2+\Pi_1)^2}
   -\frac{2(k_\perp^2)}{(\omega_n^2+{\mathbf{k}}^2+m^2+\Pi_1)^3}
   \right]\right\}\, .
   \label{V1ap}
\ee
\end{widetext}
The first term in Eq.~(\ref{V1ap}) with $\Pi_1=0$ represents the
lowest order contribution to the effective potential at finite
temperature and zero external magnetic field, usually referred to as
the {\it ideal gas} contribution~\cite{LeBellac}. In order to keep
track of the lowest order corrections in $\lambda$, we set  $\Pi_1=0$
in Eq.~(\ref{V1ap}). Thus, for the hierarchy of
energy scales considered here and dropping out the zero point energy,
the ideal gas contribution is given by~\cite{Dolan}
\begin{widetext}
\be
   \frac{T}{2}\sum_n\int\frac{d^3k}{(2\pi)^3}
   \ln(\omega_n^2+{\mathbf{k}}^2+m^2)\simeq-\frac{\pi^2T^4}{90}
   +\frac{m^2T^2}{24}-\frac{m^3T}{12\pi}-\frac{m^4}{32\pi^2}
   \ln\left(\frac{m}{4\pi T}\right)+{\mathcal{O}}(m^4)\, .
   \label{idealgas}
\ee
\end{widetext}
The second, $B$-dependent term in Eq.~(\ref{V1ap}), could potentially
ruin the nice physical picture where for weak magnetic fields, the
corrections to the ideal gas 
contribution should be proportional to a power of the parameter
$\lambda$. Fortunately it is easy to check, as we show in the
appendix, that this is not the case
as the term proportional to $(eB)^2$ in Eq.~(\ref{V1ap}) vanishes
identically. Therefore, to one-loop order, the
effective potential in the weak field case is independent of $eB$ and
is given by Eq.~(\ref{idealgas}).

Last, we compute the contribution from the ring diagrams to
the effective potential, namely, Eq.~(\ref{Vring}). Notice that for
$B=0$, it is well known that the next order correction in $\lambda$
stems from the $n=0$ term in the sum over Matsubara frequencies and is
given explicitly by~\cite{Carrington, LeBellac}.
\be
   V^{(\mbox{\tiny{ring}})}_{B=0}\!\!=\!\!-\frac{T}{12\pi}
   \left[(m^2+\Pi_1^{B=0})^{3/2} - m^3
   \right]\!\!.
   \label{VringB0}
\ee
For $eB\neq 0$, since the structure of the integrals is similar to the
case $eB=0$, the next order corrections also come from the $n=0$
term in the sum over Matsubara frequencies. To work out this case, we
notice that we can explicitly write to lowest order in $(eB)^2$
\begin{widetext}
\be
   \ln [1+\Pi^B\Delta^B (k)]&\simeq&
   \ln \left[1+\frac{\Pi_1}{\omega_n^2+{\mathbf{k}}^2+m^2+\Pi_1}\right] +
   \ln \left[1-\frac{\Pi_1(eB)^2/(8\pi T(m^2+\Pi_1)^{3/2})}
   {\omega_n^2+{\mathbf{k}}^2+m^2+\Pi_1}\right]\nonumber\\
   &+&
   \ln \left[1-\frac{\Pi_1(eB)^2}{\omega_n^2+{\mathbf{k}}^2+m^2+\Pi_1}
   \left(\frac{1}{(\omega_n^2+{\mathbf{k}}^2+m^2+\Pi_1)^2}
   -\frac{2k_\perp^2}{(\omega_n^2+{\mathbf{k}}^2+m^2+\Pi_1)^3}
   \right)\right]\, ,
   \label{desclog}
\ee
\end{widetext}
Thus, from Eq.~(\ref{Vring}), the contribution from the ring diagrams
to the effective finite temperature potential in the presence of an
external magnetic field to lowest order in $eB$ and leading order in
$\lambda$ is given by
\begin{widetext}
\be
   V^{(\mbox{\tiny{ring}})}\!\!=\!\!-\frac{T}{12\pi}\left[\left(m^2+
   \Pi_1\right)^{3/2} - m^3 \right] 
   - \frac{(eB)^2}{4\pi}\left(\frac{\Pi_1}{48}\right)
   \left(\frac{T}{(m^2+\Pi_1)^{3/2}}\right)\, ,
   \label{VringBno0}
\ee
\end{widetext}
where we have discarded a $T$ and $m$-independent infinity. Notice
that Eq.~(\ref{VringBno0}) reduces to Eq.~(\ref{VringB0}) when
$eB=0$. Also worth to note is the fact that at this order of
approximation, the corrections introduced by the ring diagrams involve
the combination $m^2+\Pi_1$, namely, the {\it thermal mass}
squared of the boson. This dependence is important since when studying
symmetry restoration, $m^2$ can vanish or even become negative. In the
former case, the assumption that $eB \ll m^2 + \Pi_1$, implied in the
calculations in Sec.~\ref{secIIIa} extended to include thermal effects,
can be satisfied as long as $eB \ll \Pi_1$. In the latter, when the mass
is corrected by thermal effects, there will be a window of temperatures
for which the combination $m^2+\Pi_1$ is non-negative. This last point
will be discussed further in section~\ref{secIV}.

\subsection{Strong field limit}\label{secIVb}

Although in the case of the EWPT, the relevant situation corresponds to
the weak field limit, for completeness of this work,
we proceed to discuss the strong field limit.

As in the previous section, we start by computing the expression for
the one-loop self-energy, this time considering the hierarchy of scales
as 
\be
   m^2 \ll T^2 \ll eB\, .
   \label{newhirarchy}
\ee
Using Eqs.~(\ref{LLL}) and~(\ref{delta})
into Eq.~(\ref{Pi}), we have to explicitly evaluate
\be
   \Pi^B_1=\lambda\frac{T}{2}\sum_n\int
   \frac{d^3k}{(2\pi)^3}
   \frac{2e^{-\frac{k_\perp^2}{eB}}}{\omega_n^2+k_3^2+eB+m^2}\, .
   \label{pilargeB}
\ee
We first perform the sum over Matsubara frequencies and the
integration over the transverse momentum. Ignoring the zero-point
energy, the result is~\cite{LeBellac} 
\be
   T\sum_n\int\frac{d^2k_\perp}{(2\pi)^2}
   \frac{2e^{-\frac{k_\perp^2}{eB}}}{\omega_n^2+\omega_0^2}=
   \left(\frac{eB}{2\pi}\right)\frac{n(\omega_0)}{\omega_0}\, ,
   \label{sumfreq}
\ee
where 
\be
   n(x)&=&\frac{1}{e^{x/T}-1}\nonumber\\
   \omega_0&=&\sqrt{k_3^2+eB+m^2}
   \label{BE}
\ee
are the Bose-Einstein thermal distribution and energy in the lowest
Landau level, respectively. Therefore, using Eq.~(\ref{sumfreq}), the
self-energy becomes
\be
   \Pi^B_1=\left(\frac{\lambda eB}{4\pi^2}\right)
   \int_{-\infty}^\infty
   dk_3\frac{n(\sqrt{k_3^2+a_0^2})}{\sqrt{k_3^2+a_0^2}}\, ,
   \label{intk3}
\ee
where $a_0^2=eB+m^2$. To carry out the integration in
Eq.~(\ref{intk3}), let us expand the distribution function in terms of
a geometric series. Thus, after the exchange of the sum and the
integral, we get 
\be
   \Pi^B_1&=&2\left(\frac{\lambda eB}{4\pi^2}\right)
   \sum_{l=1}^\infty\int_0^\infty
   dk_3\frac{e^{-l(\sqrt{k_3^2+a_0^2})/T}}{\sqrt{k_3^2+a_0^2}}
   \nonumber\\
   &=&\left(\frac{\lambda eB}{2\pi^2}\right)
   \sum_{l=1}^\infty K_0\left(\frac{la_0}{T}\right)\, ,
   \label{sumint}
\ee
where $K_0$ is the modified Bessel function of order $0$. For $a_0\gg T$
the largest contribution in Eq.~(\ref{sumint}) comes from the term with
$l=1$. Thus, from the asymptotic expansion of $K_0(z)$ we obtain 
\be
   \Pi^B_1&\stackrel{eB\rightarrow \infty}{\longrightarrow}&
   \frac{\lambda
   eB}{(2\pi)^{3/2}}\left(\frac{T^2}{eB+m^2}\right)^{1/4}\!\!
   e^{-\sqrt{eB+m^2}/T}.
   \label{pi1largeB}
\ee
Notice that Eq.~(\ref{pi1largeB}) means that the self energy in the
strong $eB$ limit is exponentially small, which means that the
contribution from the ring diagrams is negligible.

Next, we proceed to the computation of the effective potential to
one-loop order, Eq.~(\ref{V1}). Notice that for large $eB$, we can
write Eq.~(\ref{LLL}) as 
\be
   \Delta^B(k)&\simeq&
   \left(\frac{2}{eB}\right)\frac{1-\frac{k_\perp^2}{eB}}
   {(1+\frac{\omega_n^2+k_3^2+m^2}{eB})}\nonumber\\
   &=&\left(\frac{2}{eB}\right)\frac{1}
   {\left(1+\frac{k_\perp^2}{eB}\right)
   \left(1+\frac{\omega_n^2+k_3^2+m^2}{eB}\right)}\nonumber\\
   &\simeq&
   \frac{2}{(\omega_n^2+{\mathbf{k}}^2+m^2+eB)}\, .
   \label{DeBlar}
\ee
Therefore, the integral at hand can be written as
\begin{widetext}
\be
   \frac{T}{2}\sum_n\int\frac{d^3k}{(2\pi)^3}\ln
   [\Delta^B (k)]^{-1}=\frac{T}{2}\sum_n\int\frac{d^3k}{(2\pi)^3}
   \ln\left(\frac{\omega_n^2+\omega^2}{2}\right)\, ,
   \label{U}
\ee
\end{widetext}
where $\omega^2\equiv{\mathbf{k}}^2+m^2+eB$. Equation~(\ref{U})
represents the ideal gas contribution. In this case however, this
contribution depends on $eB$ through $\omega$. 

To evaluate the right-hand side of Eq.~(\ref{U}), we can take the
derivative with respect to $\omega$, perform the sum and integrate
again with respect to $\omega$~\cite{LeBellac}. The result is 
\be
   \frac{T}{2}\sum_{n}&&\!\!\!\!\!\!\int\frac{d^3k}{(2\pi)^3}\ln
   [\Delta^B (k)]^{-1}=\nonumber\\
   \frac{1}{2}&&\!\!\!\!\!\!\int\frac{d^3k}{(2\pi)^3}
   \left\{\omega + 2T\ln(1-e^{\omega/T}) + \alpha\right\}\, ,
   \label{derint}
\ee
where $\alpha$ is a constant independent of $eB$ and $T$ and can thus
be ignored. The term proportional to $\omega$ in Eq.~(\ref{derint})
gives rise to a temperature independent, though $eB$-dependent
infinity and corresponds to the zero-point energy, which has been
already ignored to deduce Eq.~(\ref{sumfreq}) and we also ignore
here. The procedure can be put in more elegant terms by defining a
renormalized effective potential subtracting the value of this a
$T=0$. Since this discussion is standard (see for example
Ref.~\cite{LeBellac}), we omit it here for the ease of the discussion
and therefore take
\be
   \!\!\frac{T}{2}\sum_{n}\!\!\int\!\!\frac{d^3k}{(2\pi)^3}\ln
   [\Delta^B (k)]^{-1}=\!\!
   T\!\!\int\!\!\frac{d^3k}{(2\pi)^3}
   \ln(1-e^{\omega/T}).
   \label{nozero}
\ee
For large $eB$, we can approximate the integral in the right-hand side
of Eq.~(\ref{nozero}) by
\be
   -T\int\frac{d^3k}{(2\pi)^3}e^{-\omega/T}
   &=&-\left[\frac{T^2(eB+m^2)}{2\pi^2}\right]\nonumber\\
   &\times&
   K_2\left(\frac{\sqrt{eB+m^2}}{T}\right)\!\!,
   \label{casifinlargeeB}
\ee
where $K_2$ is the modified Bessel function of order $2$. From the
asymptotic expansion of $K_2$, we finally get
\be
   V^{(1)}\stackrel{eB\rightarrow \infty}{\longrightarrow}
   -\left[\frac{T^5(eB+m^2)^{3/2}}{(2\pi)^3}\right]^{1/2}
   e^{-\sqrt{eB+m^2}/T}.
   \label{finlargeeB}
\ee
We now proceed to discuss symmetry restoration at finite
temperature. For the analysis, we restrict ourselves to the weak
field limit which, as previously indicated, is the relevant scenario for
the description of the EWPT.

\section{Symmetry restoration}\label{secIV}

To address symmetry restoration, it is convenient to write down the
explicit model for the theory. We will consider the Lagrangian
\be
   {\mathcal{L}}=(D_\mu\phi)^\dagger D^\mu\phi +\mu^2\phi^\dagger\phi -
   \frac{\lambda}{4}(\phi^\dagger\phi)^2 - \frac{1}{4}F_{\mu\nu}
   F^{\mu\nu}\, ,
   \label{AbelianHiggs}
\ee
where
\be
   D_\mu\phi&=&\partial_\mu\phi - ieA_\mu\phi\nonumber\\
   F_{\mu\nu}&=&\partial_\mu A_\nu-\partial_\nu A_\nu\, ,
   \label{defsDF}
\ee
and
\be
   \mu^2 ,\ \lambda > 0\, .
   \label{mulambda}
\ee
The Lagrangian in Eq.~(\ref{AbelianHiggs}) represents the interaction
of a charged scalar field with an electromagnetic field and
is commonly known as the Abelian-Higgs model. We take $F_{\mu\nu}$ as
the external electromagnetic field containing only the magnetic
component. It is well known that 
for this model, with a local, spontaneously broken gauge symmetry, the
gauge field $A_\mu$ acquires a finite mass and thus cannot represent
the physical situation of a massless photon interacting with the
charged scalar $\phi$. However, since the physically interesting
situation to what the findings of this work will apply is the SM, with
an $U(1)\times SU(2)$ broken gauge symmetry, where the Higgs
mechanisms ensures that the photon remains massless, for the discussion
we will ignore the mass generated for $A_\mu$ and will concentrate on
the scalar sector. 

\begin{figure}[t!] 
\vspace{0.4cm}
{\centering
\resizebox*{0.4\textwidth}
{0.2\textheight}{\includegraphics{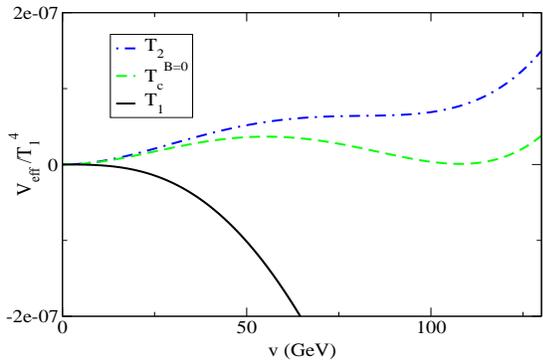}}
\par}
\caption{Finite temperature effective potential for the case $eB=0$
for three different temperatures. We use the values of the parameters
$\mu=20$ GeV and $\lambda=0.0025$.} 
\label{fig5}
\end{figure}

The complex fields $\phi$ and $\phi^\dagger$ can be equivalently
expressed in terms of the two Hermitian fields $\sigma$ and $\chi$ by
means of the definition
\be
   \phi (x)&=&\frac{1}{\sqrt{2}}[\sigma (x) +i\chi (x)]\nonumber\\
   \phi^\dagger (x)&=&\frac{1}{\sqrt{2}}[\sigma (x) -i\chi (x)]\, .
   \label{defssc}
\ee

The Lagrangian in Eq.~(\ref{AbelianHiggs}) is symmetric under the
transformation $\phi\rightarrow -\phi$, however, the vacuum is
not. Selecting the vacuum about which perturbative calculations can be
performed, we shift the field by its classical value $v$ writing
\be
   \sigma \rightarrow v + \sigma\, .
   \label{shift}
\ee
After the shift, the mass of the fields $\sigma$ and $\chi $ become
\be
   m^2_1(v)&=&\frac{3}{4}\lambda v^2 - \mu^2\nonumber\\
   m^2_2(v)&=&\frac{1}{4}\lambda v^2 - \mu^2\, ,
   \label{mass}
\ee
\\
respectively. To lowest order (tree level) the potential is
\be
   V^{({\mbox{\tiny{tree}}})}=-\frac{1}{2}\mu^2v^2 + 
   \frac{1}{16}\lambda v^4\, .
   \label{tree}
\ee
\\
To next order, we should include the zero-temperature part of the
one-loop potential, given by
\be
   V^{(1)}_{\mbox{\tiny{vac}}}=\frac{1}{2}\int\frac{d^3k}{(2\pi)^3}
   \left(\sqrt{{\mathbf{k}}^2+m_1^2} + \sqrt{{\mathbf{k}}^2+m_2^2}
   \right)\, .
   \label{V1vac}
\ee
The integral in Eq.~(\ref{V1vac}) is divergent and the theory must be
renormalized. This can be achieved by introducing counter-terms in the
Lagrangian of the form
\be
   {\mathcal{L}}_{\mbox{\tiny{ct}}}=\frac{A}{2}v^2 + \frac{B}{16}v^4 +
   C\, ,
   \label{Lct}
\ee
where $C$ is a constant that can be used to cancel the $v$-independent
part of the vacuum energy and $A$ and $B$ are determined by requiring
that the infinities cancel. By this means, the effective potential
up to one loop is~\cite{Carrington}
\be
   V_{\mbox{\tiny{vac}}}=
    -\frac{1}{2}\mu^2v^2 + 
   \frac{\lambda}{16}v^4 +
   \frac{1}{32\pi^2}\sum_i  m_i^4\ln (\frac{m_i}{2}) \, ,
   \label{V1vacexpl}
\ee
where $m_i$, $i=1,2$ are given by Eqs.~(\ref{mass}). 

In the weak field limit, the finite-temperature effective potential,
up to the contribution from the ring diagrams, is
given by adding Eqs.~(\ref{idealgas}) and~(\ref{VringBno0}) to
Eq.~(\ref{V1vacexpl}), accounting for the contributions from the two
fields $\sigma$ and $\chi$. Dropping the $v$-independent term, the
result is 
\begin{widetext}
\be
   V(v)=-\frac{1}{2}\mu^2v^2
   +\frac{1}{16}\lambda v^4
   +\sum_i\left[\frac{m_i^2T^2}{24}
   -\frac{T}{12\pi}\left(m_i^2 +
   \Pi_1
   \right)^{3/2}-
   \frac{(eB)^2}{4\pi}\left(\frac{\Pi_1}{48}\right)
   \left(\frac{T}{(m_i^2+\Pi_1)^{3/2}}\right)
   +{\mathcal {O}}(m_i^4)\right]
   \label{Vofv}
\ee
\end{widetext}
where $m_i$, $i=1,2$ are given by Eqs.~(\ref{mass}) and $\Pi_1$ is
given by Eq.~(\ref{Pi1}). Notice that the terms proportional to $Tm_i^3$ 
and to $\ln(m_i)$ have offset each other when adding up all the
contributions. Also, in order for the terms involving the square root
of the boson's thermal mass to be real, the temperature must be such
that 
\be
   T>T_1\equiv \mu\sqrt\frac{12}{\lambda}\, ,
   \label{T1}
\ee
which defines a lower bound for the temperature. Notice that the
development of an imaginary part in the effective potential signals
the onset of {\it spinodal decomposition} and the 
pase transition is quickly completed. This happens when the
combination $m^2 + \Pi_1$ becomes negative. For all values of $v$,
this occurs for temperatures lower than $T_1$ defined in
Eq.~(\ref{T1}).  

\begin{figure}[t!] 
\vspace{0.4cm}
{\centering
\resizebox*{0.4\textwidth}
{0.2\textheight}{\includegraphics{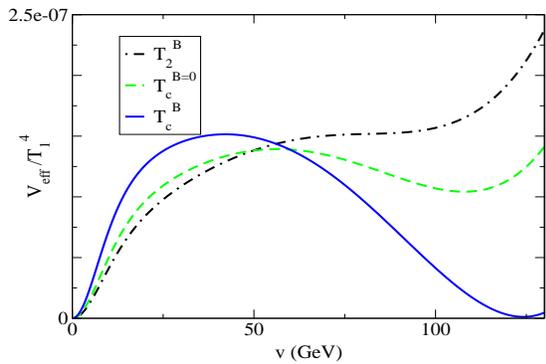}}
\par}
\caption{Finite temperature effective potential for the case $eB\neq 0$
for three different temperatures. We use the values of the parameters
$\mu=20$ GeV, $\lambda=0.0025$ and $e=0.3$ and parametrize the
magnetic field strength as $B=b(100)^2$ (GeV)$^2$, using
$b=0.01$.} 
\label{fig6}
\end{figure}

Figure~\ref{fig5} shows the finite temperature effective potential,
discarding $v$-independent terms, for
the case $B=0$ for three different temperatures, the above defined
$T_1$, a temperature $T_c^{B=0}>T_1$ where the two minima coincide and a
temperature $T_2>T_c^{B=0}$ where the second minimum of the potential
disappears. For the calculation, we have used $\mu=20$ GeV
and $\lambda=0.0025$.

Figure~\ref{fig6} shows the finite temperature effective potential,
discarding $v$-independent terms, for 
the case $B\neq 0$ for the above defined temperature $T_c^{B=0}$ and two
more values of the temperature: the curve with degenerate minima
corresponds to a temperature $T_c^B<T_c^{B=0}$ and the curve where the
second minimum has disappeared corresponds to a temperature
$T_2^B>T_c^{B=0}$. For the calculation, we have used the same values of the
parameters as for the case with $B=0$, taking $e=0.3$ and have
parametrized the magnetic field strength as $B=b(100)^2$ (GeV)$^2$,
using $b=0.01$. 

\begin{figure}[t!] 
\vspace{0.4cm}
{\centering
\resizebox*{0.4\textwidth}
{0.2\textheight}{\includegraphics{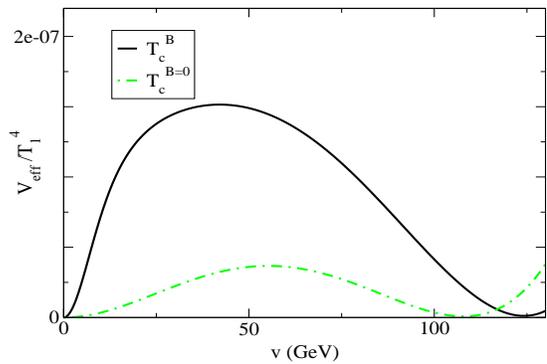}}
\par}
\caption{Comparison between the finite temperature effective potential
for the cases with and without magnetic field evaluated at the
temperatures where the minima of the potentials are degenerate. The
height of the barrier between minima is higher in the case $eB\neq 0$,
signaling a stronger first order phase transition. We use
the values of the parameters $\mu=20$ GeV, $e=0.3$ and
$\lambda=0.0025$ and parametrize the magnetic field strength as
$B=b(100)^2$ (GeV)$^2$, using $b=0.01$.} 
\label{fig7}
\end{figure}

Notice that the effect of the magnetic field is twofold:
first it delays the starting of the phase transition down to a
temperature $T_c^B<T_c^{B=0}$ and second, it makes the transition strongly
first order. This second feature is best seen in Fig.~\ref{fig7}
where we compare the effective potential obtained for $B=0$ and 
$B\neq 0$, discarding $v$-independent terms, for the two temperatures
$T_c^{B=0}$ and $T_c^B$. For the latter 
temperature, the height of the barrier becomes larger signaling a
stronger first order phase transition as compared to the case with
$B=0$. The origin of this feature is that the corrections introduced
by the magnetic field are inversely proportional to a power of the
boson's thermal mass and thus are larger for the value of $v$ when the
mass parameters of Eqs.~(\ref{mass}) vanish.

\section{Conclusions}\label{secV}

In this work we have studied the effects that the presence of a
constant external magnetic field has on the effective potential for a
charged scalar at finite temperature, up to the contribution from
the ring diagrams. By studying symmetry restoration in the weak field
limit we have found that the magnetic field is able to produce a
stronger first order phase transition signaled by an increase in the
height of the barrier between degenerate minima with respect to the
case without magnetic field. The temperature for the onset of the
phase transition is also lowered by the presence of the magnetic field
as compared to the case without magnetic field.

The findings of this work show that there exists room in the parameter
space of this theory where the effects of the magnetic field could
be important. In particular, an extension of these ideas to the case
of the SM degrees of freedom to describe the EWPT could be significant
for the problem of the generation of baryon number. This is work under
progress and will be reported as a sequel of the present one. 

\section*{Acknowledgments}

A.A. wishes to thank J. Magnin for his hospitality during a visit to
CBPF where part of this work was completed. G.P. wishes to thank
IA-UNAM for their hospitality during the completion of this
work. Support for this work has 
been received in part by DGAPA-UNAM under PAPIIT grant number IN108001
and by CONACyT-M\'exico under grant number 40025-F.  

\section*{Appendix}

We start first by evaluating the integrals
\be
   J_1&=&T\sum_n\int\frac{d^3k}{(2\pi)^3}
   \frac{1}{(\omega_n^2+{\mathbf{k}}^2+m^2)^2}\nonumber\\
   J_2&=&\left(\frac{4}{3}\right)T\sum_n\int\frac{d^3k}{(2\pi)^3}
   \frac{\mathbf{k}^2}{(\omega_n^2+{\mathbf{k}}^2+m^2)^3}\, ,
   \label{J1J2}
\ee
appearing in the computation of the effective potential in the weak
field limit, Eq.~(\ref{V1ap}). To this end, notice that we can write
\be
   J_1&=&T\sum_n\int\frac{d^3k}{(2\pi)^3}f({\mathbf{k}}^2)\nonumber\\
   J_2&=&-\left(\frac{1}{3}\right)T\sum_n\int\frac{d^3k}{(2\pi)^3}
   {\mathbf{k}}\cdot{\mathbf{\nabla}}_kf({\mathbf{k}}^2)\, ,
   \label{nabla}
\ee
where
\be
   f({\mathbf{k}}^2)=\frac{1}{(\omega_n^2+{\mathbf{k}}^2+m^2)^2}\, .
   \label{deff}
\ee
Integrating by parts the second of Eqs.~(\ref{nabla}) we get
\be
   J_2&=&-\left(\frac{1}{3}\right)T\sum_n\int\frac{d^3k}{(2\pi)^3}
   {\mathbf{\nabla}}_k\cdot\left({\mathbf{k}}f({\mathbf{k}}^2)\right)
   \nonumber\\
   &+&\left(\frac{1}{3}\right)T\sum_n\int\frac{d^3k}{(2\pi)^3}
   f({\mathbf{k}}^2){\mathbf{\nabla}}_k\cdot{\mathbf{k}}\, .
   \label{intpart}
\ee
The first of the terms in Eq.~(\ref{intpart}) can be converted into a
surface integral and since $f({\mathbf{k}}^2)$ decreases faster than
${\mathbf{k}}^{-2}$ at infinity, this surface term vanishes. Also,
using that ${\mathbf{\nabla}}_k\cdot{\mathbf{k}}=3$, we finally get
$J_1=J_2$.

\end{document}